\documentclass[9pt,twocolumn,twoside]{osajnl}

\usepackage{upgreek}
\usepackage{graphicx}
\usepackage{subcaption}
\usepackage{float}
\usepackage[export]{adjustbox}
\usepackage{dblfloatfix} 
\usepackage{siunitx}

\journal{ol} 

\setboolean{shortarticle}{true} 

\title{InP femtosecond mode-locked laser in compound feedback cavity with switchable repetition rate}

\author[1,*]{Mu-Chieh Lo}
\author[1]{Robinson Guzm\'an}
\author[1]{Guillermo Carpintero}

\affil[1]{Departamento de Tecnología Electr\'onica, Universidad Carlos III de Madrid, Av de la Universidad, 30. 28911 Leganés, Madrid, Spain}

\affil[*]{Corresponding author: mlo@ing.uc3m.es}

\ociscodes{(140.4050) Mode-locked lasers; (140.5960) Semiconductor lasers; (250.5300) Photonic integrated circuits.}

\begin{abstract}
A monolithically integrated mode-locked semiconductor laser is proposed. The compound ring cavity is composed of a colliding pulse mode locking subcavity and a passive Fabry-Pérot feedback subcavity. These two 1.6-mm-long subcavities are coupled by using on-chip reflectors at both ends enabling harmonic mode locking. By changing DC-bias conditions, optical mode spacing from 50 up to 450 GHz are experimentally demonstrated. Ultrafast pulses shorter than 0.3 ps emitted from this laser diode are shown in autocorrelation traces.
\end{abstract}

\setboolean{displaycopyright}{true}

\begin{document}

\maketitle


Passively mode-locked lasers (MLL) in photonic integrated circuits (PIC) at millimeter-wave (mmW) and terahertz (THz) rates have attracted increasing attention in recent years for communications, spectroscopy, and microwave photonics \cite{Delfyett,Nagatsuma}. To increase the repetition rate of MLL above intrinsic cavity round-trip frequency, harmonic mode locking (HML) schemes have been reported where multiple pulses are produced in one single round trip \cite{Marsh}. HML can be implemented by means of commonly adopted colliding pulse ML \cite{Chen, Martins-Filho, Shimizu}, coupled cavity ML \cite{Avrutin,Yanson,Thiessen}, methods based on the selectivity of harmonic orders owing to the spectral filtering of distributed Bragg reflector (DBR) \cite{Arahira,Hou:17}, and the temporal multiplexing of Mach-Zehnder inteference (MZI) interleaver \cite{Lo}. However, these harmonic-selectable structures depend on either multiple segments or sub-wavelength grating components which are difficult to design, manufacture or utilize.

\begin{figure}[htbp]
\centering
    \begin{subfigure}[t]{\linewidth}
    \centering
    \captionsetup{justification=centering}
    \caption[c]{}
    \centering
    \includegraphics[scale=0.12]{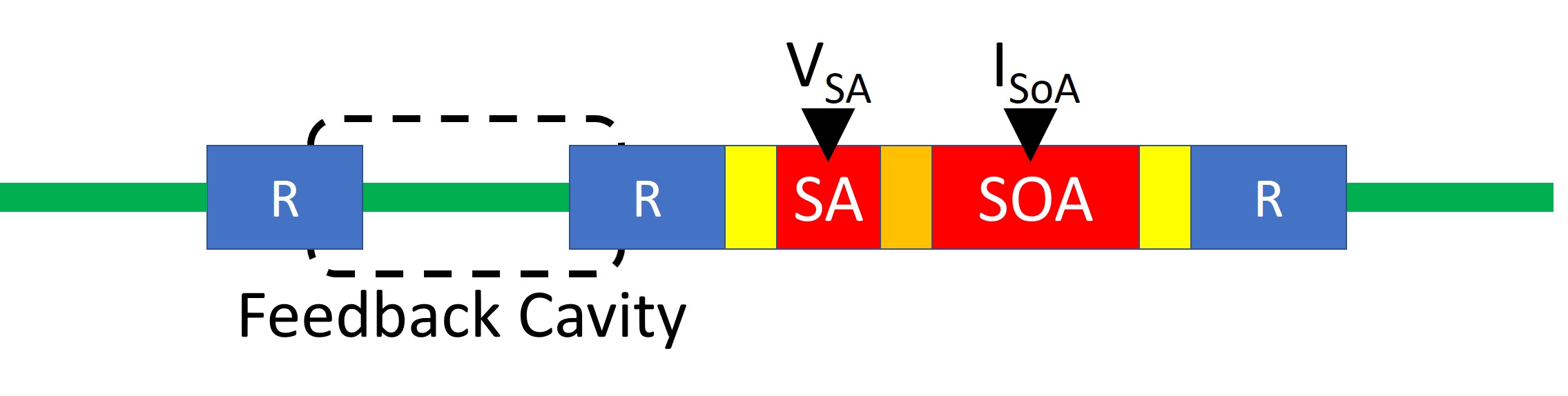}
    \end{subfigure}
    \begin{subfigure}[t]{\linewidth}
    \centering
    \captionsetup{justification=centering}
    \caption[c]{}
    \centering
    \includegraphics[scale=0.12]{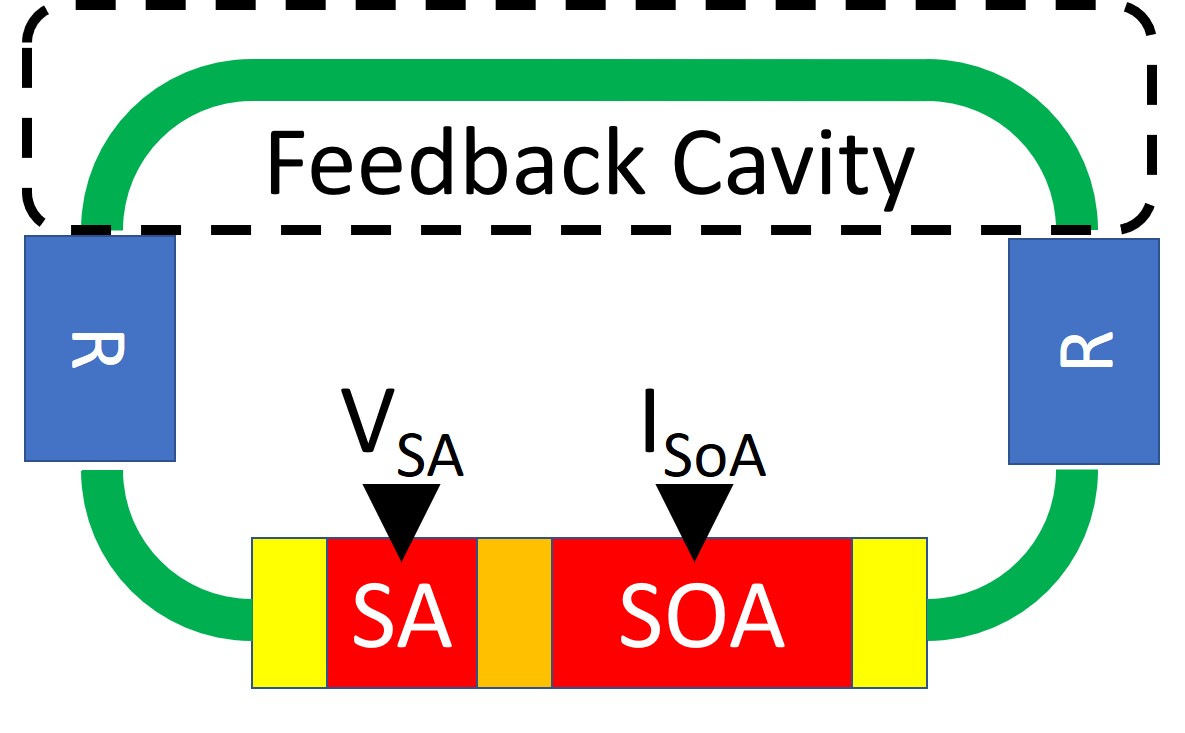}
    \end{subfigure}
\caption{(a) Two-section mode-locked laser coupled with feedback cavity. (b) Two-section mode-locked laser coupled with compound feedback cavity.}
\label{fig:introduction}
\end{figure}

\begin{figure*}[!htbp]
\centering
  \begin{subfigure}[b]{.33\textwidth}
    \centering
        \captionsetup{justification=centering}
        \caption[c]{}\label{fig:2a}
            \centering
    \includegraphics[width=.92\textwidth]{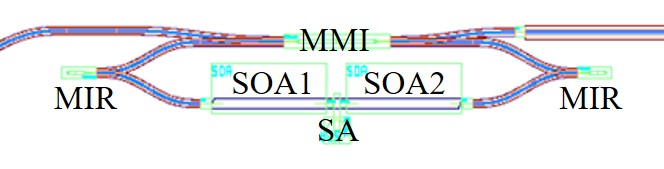}
  \end{subfigure}%
  \begin{subfigure}[b]{.33\textwidth}
    \centering
        \captionsetup{justification=centering}
        \caption[c]{}\label{fig:2b}
            \centering
    \includegraphics[width=.92\textwidth]{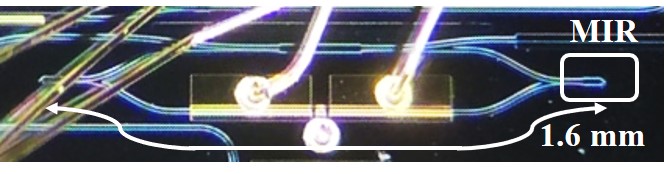}
  \end{subfigure}%
  \begin{subfigure}[b]{.33\textwidth}
    \centering
        \captionsetup{justification=centering}
        \caption[c]{}\label{fig:2c}
            \centering
    \includegraphics[width=.87\textwidth]{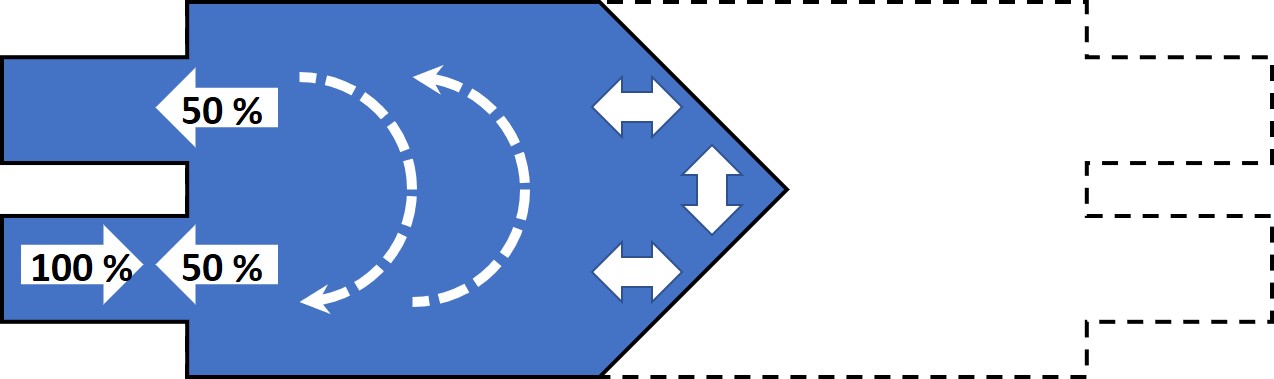}
  \end{subfigure}\\%
  \caption{(a) Layout of the PIC under test. (b) Micrograph of the PIC under test. (c) Schematic of multimode interference reflector (MIR).}\label{fig:2}
\end{figure*}

In this paper, we propose a simple monolithic MLL featuring switchable order of harmonic. The laser cavity consists of an active linear mode locking (ML) subcavity, coupled with another passive linear subcavity via two foundry-provided reflective coupler building blocks. These two linear 25-GHz subcavities form an end-to-end loop by folding the additional coupled cavity \cite{Avrutin} which allows for HML operations. This PIC was developed in a generic foundry platform \cite{Smit, Augustin}. The recorded optical frequency combs with mode spacing of 50 GHz and 100 GHz are presented. Mode locking is exhibited through autocorrelation traces with pulse duration faster than 300 femtoseconds. Harmonics up to 450 GHz (18th) are observed.


Fig. 1(a) presents the classic two-section mode-locked semiconductor laser comprising a gain section (SOA) and a saturable absorber (SA), integrated with optical feedback subcavity, as reported in \cite{Avrutin} where HML was seen experimentally the first time. Those Fabry-Pérot subcavities were defined by reflectors (R) using facet edge or etched groove. In \cite{Thiessen} the design was repeated based on DBR gratings where a fraction of the beam emitted from the laser was fed back into itself via the intermediate reflector to achieve higher signal purity and stability. To further study the behavior of such an integrated feedback, the advanced coupled compound cavity is proposed as depicted in Fig. 1(b). This compound ring can be regarded as a folded version of feedback cavity design in Fig. 1(a). The two reflectors define two subcavities in which optical pulses travel, split, and recombine.

\begin{figure*}[hbp]
\centering
  \begin{subfigure}[t]{.24\textwidth}
    \centering
        \captionsetup{justification=centering}
        \caption[c]{}\label{fig:3a}
            \centering
            \begin{center}
    \includegraphics[width=.99\textwidth]{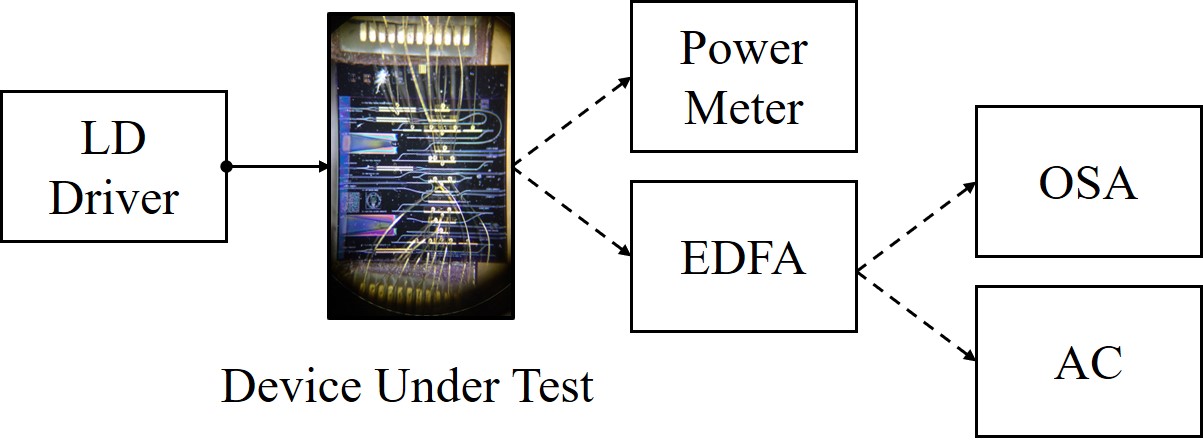}
    		\end{center}
  \end{subfigure}%
  \begin{subfigure}[t]{.24\textwidth}
    \centering
        \captionsetup{justification=centering}
        \caption[c]{}\label{fig:3b}
            \centering
    \includegraphics[width=.99\textwidth]{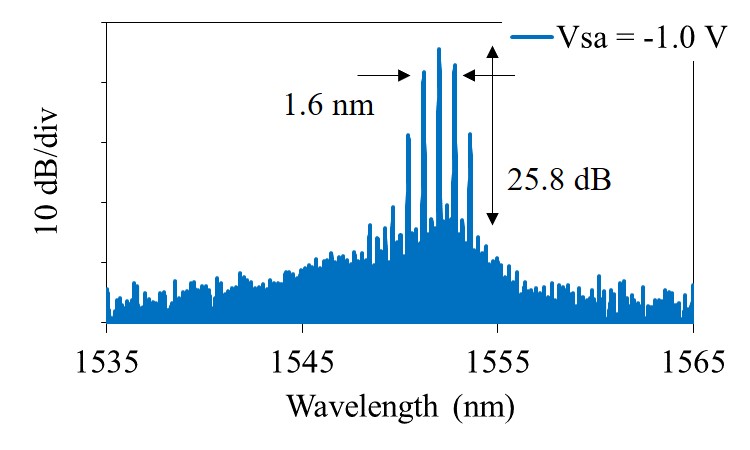}
  \end{subfigure}%
  \begin{subfigure}[t]{.24\textwidth}
    \centering
        \captionsetup{justification=centering}
        \caption[c]{}\label{fig:3c}
            \centering
    \includegraphics[width=.99\textwidth]{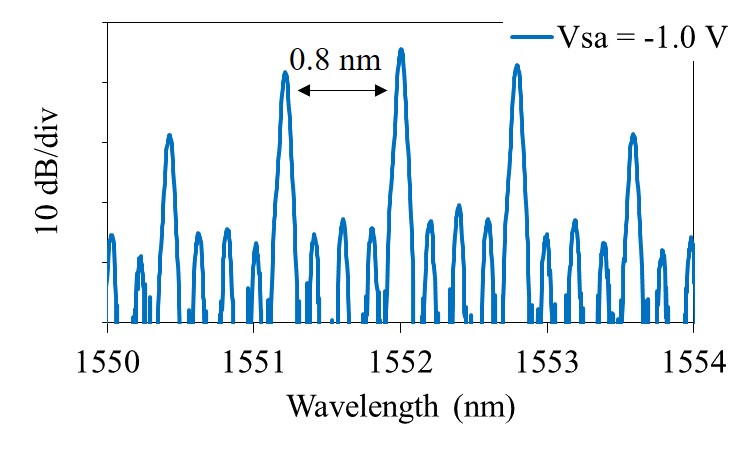}
  \end{subfigure}%
  \begin{subfigure}[t]{.24\textwidth}
    \centering
        \captionsetup{justification=centering}
        \caption[c]{}\label{fig:3d}
            \centering
    \includegraphics[width=.99\textwidth]{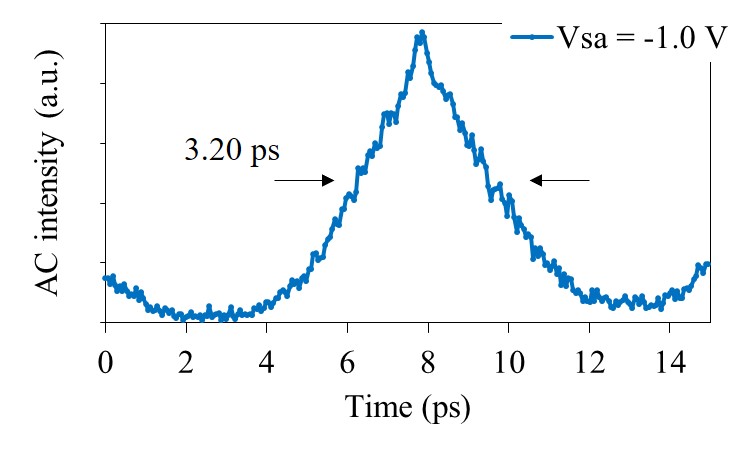}
  \end{subfigure}\\%
    \begin{subfigure}[t]{.24\textwidth}
    \centering
        \captionsetup{justification=centering}
        \caption[c]{}\label{fig:3e}
            \centering
    \includegraphics[width=.99\textwidth]{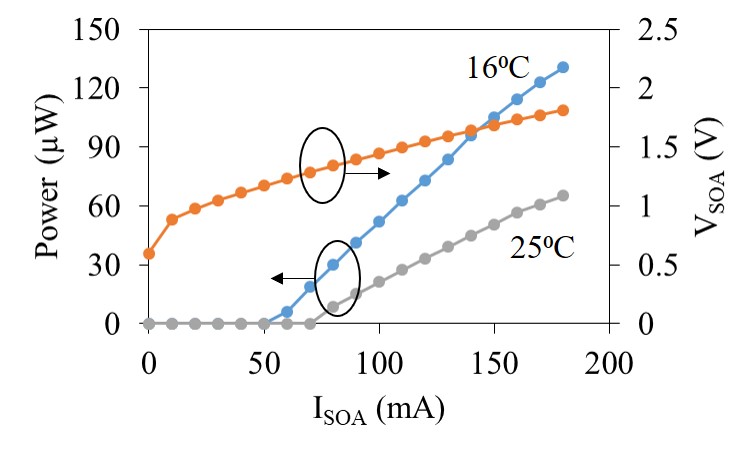}
  \end{subfigure}%
  \begin{subfigure}[t]{.24\textwidth}
    \centering
        \captionsetup{justification=centering}
        \caption{}\label{fig:3f}
            \centering
    \includegraphics[width=.99\textwidth]{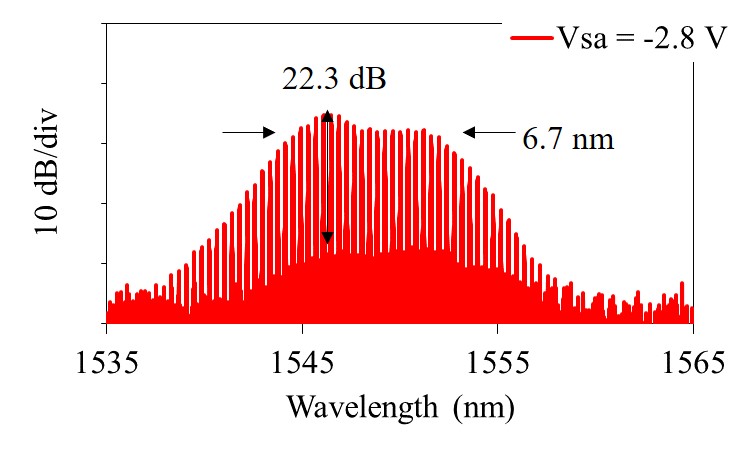}
  \end{subfigure}%
  \begin{subfigure}[t]{.24\textwidth}
    \centering
        \captionsetup{justification=centering}
        \caption{}\label{fig:3g}
            \centering
    \includegraphics[width=.99\textwidth]{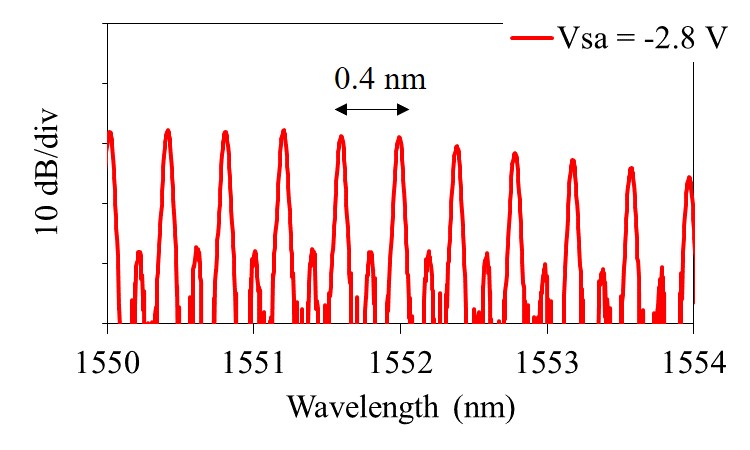}
  \end{subfigure}%
  \begin{subfigure}[t]{.24\textwidth}
    \centering
        \captionsetup{justification=centering}
        \caption{}\label{fig:3h}
            \centering
    \includegraphics[width=.99\textwidth]{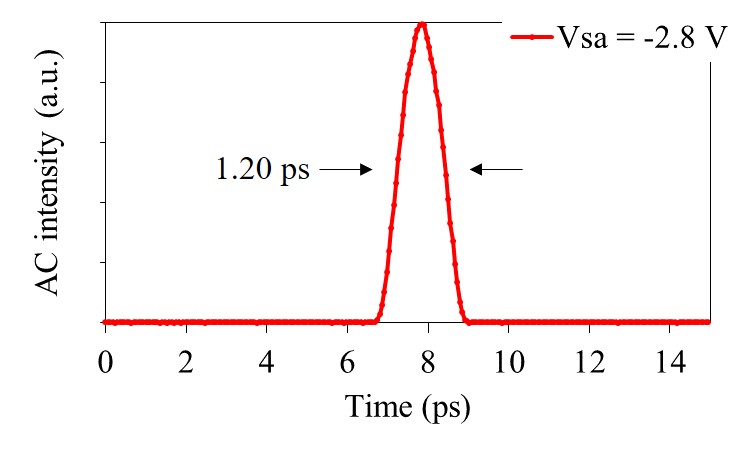}
  \end{subfigure}%
  \caption{(a) Experimental setup. (b) Optical spectrum, (c) detailed optical spectrum, and (d) AC trace under $I_{SOA}$ = 100 mA and $V_{SA}$ = -1.0 V. (e) L-I-V characteristic at \SI{16}{\celsius} and \SI{25}{\celsius}. (f) Optical spectrum,(g) detailed optical spectrum, and (h) AC trace under $I_{SOA}$ = 100 mA and $V_{SA}$ = -2.8 V.}\label{fig:3}
\end{figure*}

The mask layout and microscope photograph of realized PIC are shown in Fig. 2(a) and (b), respectively.  The MLL cavity is composed of two 1.6-mm Fabry-Pérot subcavities each corresponding to 25-GHz round-trip rate, coupled via two MIRs (Multimode Interference Reflector) building blocks \cite{Kleijn} available in the generic foundry  platform \cite{Smit,Augustin}. The lower subcavity contains a 20-$\upmu$m SA surrounded symmetrically by two 345-$\upmu$m SOAs. Both SOAs are forward-biased with current source $I_{SOA}$ and the SA is reverse-biased with voltage source $V_{SA}$. The symmetric configuration gives rise to colliding pulse ML \cite{Chen}, generating a pulse train of second harmonic, 50 GHz. The upper subcavity is a passive waveguide loaded with a 2x2 MMI (Multimode Interference) coupler. As shown in Fig. 2(c), the 2-port MIR splits the incident lightwave into reflection ($\sim$50\%) and transmission ($\sim$50\%). Split and reflected by a MIR, the generated pulse trains propagate back through the active components in the lower arm. Meanwhile, the optical pulses splitting and transmitting to the other port travel along the passive MMI-loaded feedback cavity. Consequently, the reflected and transmitted pulses merge at the other MIR, prior to the next iteration of splitting. Confined by the two MIRs, the optical pulses divide and recombine, traveling back and forth in the compound cavity. The pulses forking into the lower arm are amplified and reshaped in intensity and phase, while the others into the upper arm are attenuated and losing 3-dB due to the 2x2 MMI. The interference of these two split pulses may lead to further complex mechanism.

In the upper cavity, the 50:50 2x2 MMI guides part of the optical pulses out of the compound cavity, towards waveguide outputs at cleaved edges which are angled at 7° and applied with anti-reflection (AR) coatings to minimize facet reflections. Isolations were inserted between active components to avoid unwanted current flows. Between deeply and shallowly etched waveguides transitions were deployed to butt-joint both type waveguides. A lensed fiber was used to collect the optical signal from the waveguide output to diagnostic instruments.


\begin{figure*}[!htbp]
\centering
  \begin{subfigure}[t]{.16\textwidth}
    \centering
        \captionsetup{justification=centering}
        \caption[c]{}\label{fig:5a}
            \centering
    \includegraphics[width=.99\textwidth]{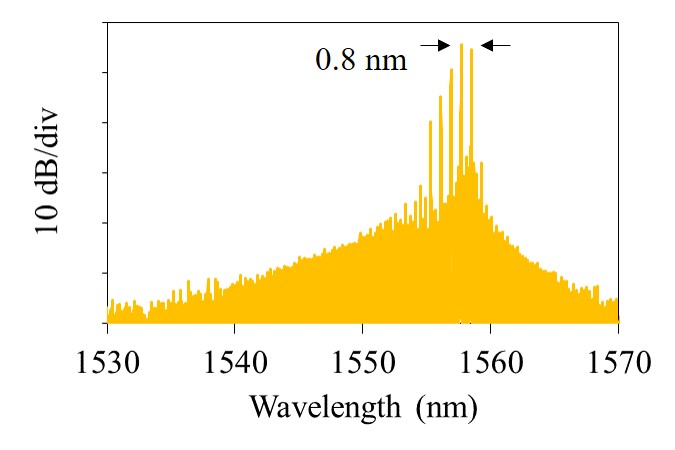}
  \end{subfigure}%
  \begin{subfigure}[t]{.16\textwidth}
    \centering
        \captionsetup{justification=centering}
        \caption[c]{}\label{fig:5b}
            \centering
    \includegraphics[width=.99\textwidth]{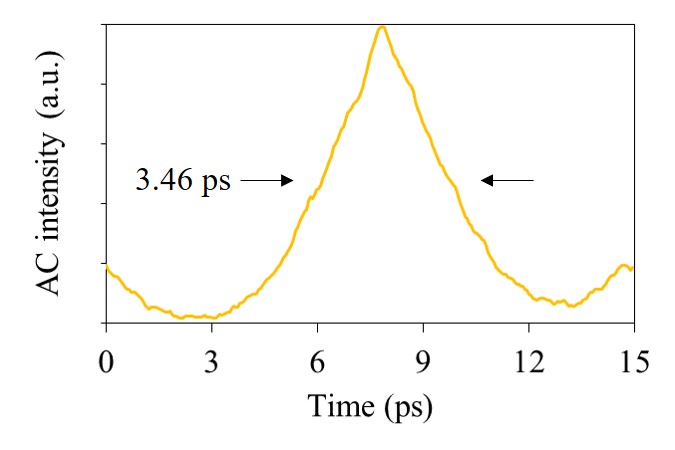}
  \end{subfigure}%
  \begin{subfigure}[t]{.16\textwidth}
    \centering
        \captionsetup{justification=centering}
        \caption[c]{}\label{fig:5d}
            \centering
    \includegraphics[width=.99\textwidth]{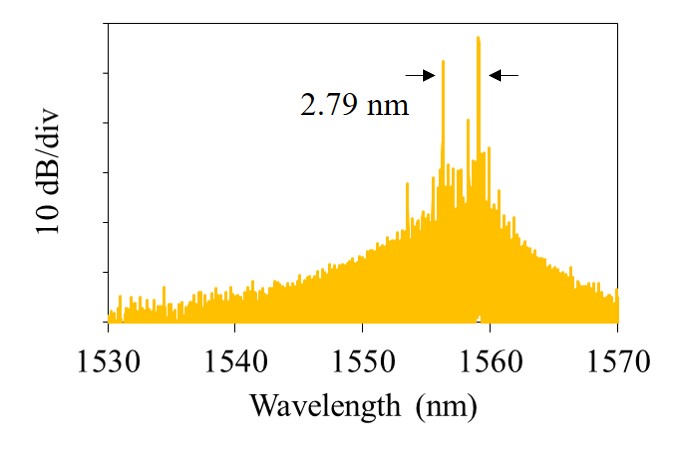}
  \end{subfigure}%
    \begin{subfigure}[t]{.16\textwidth}
    \centering
        \captionsetup{justification=centering}
        \caption[c]{}\label{fig:5e}
            \centering
    \includegraphics[width=.99\textwidth]{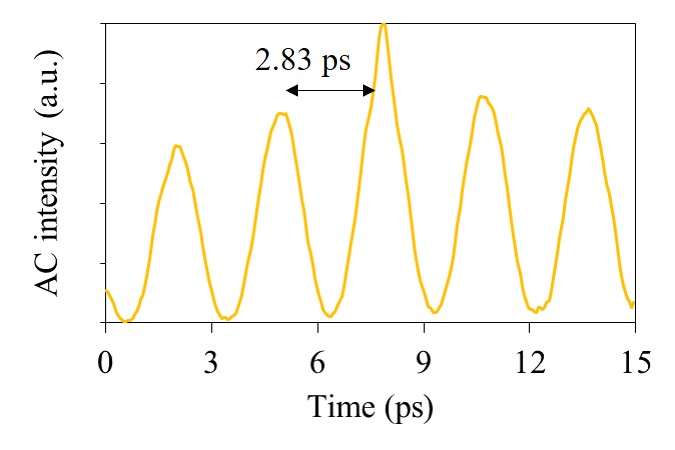}
  \end{subfigure}%
  \begin{subfigure}[t]{.16\textwidth}
    \centering
        \captionsetup{justification=centering}
        \caption{}\label{fig:5f}
            \centering
    \includegraphics[width=.99\textwidth]{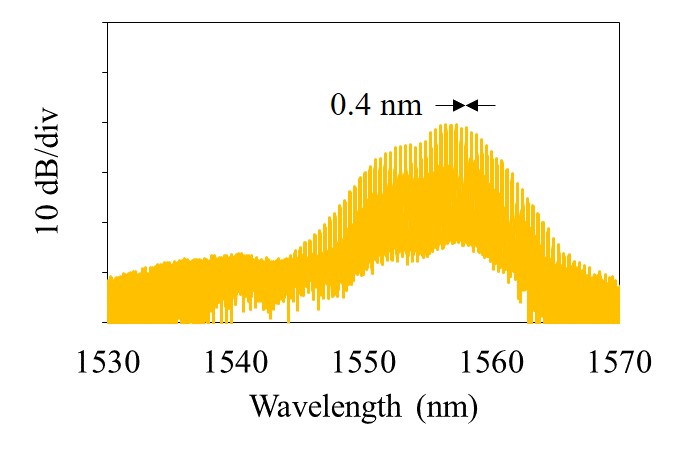}
  \end{subfigure}%
  \begin{subfigure}[t]{.16\textwidth}
    \centering
        \captionsetup{justification=centering}
        \caption{}\label{fig:5h}
            \centering
    \includegraphics[width=.99\textwidth]{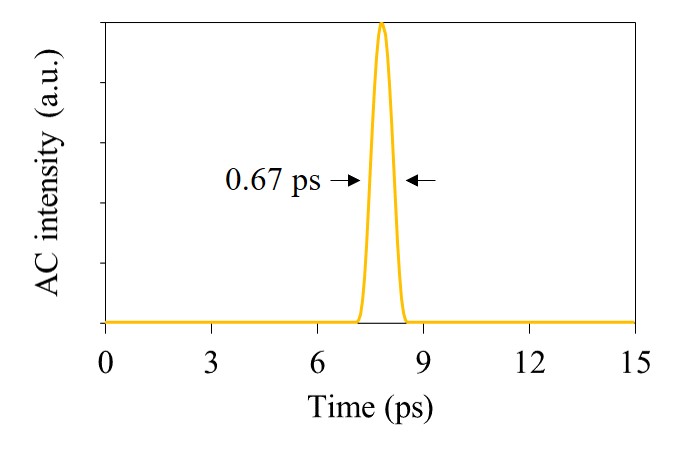}
  \end{subfigure}%
  \caption{(a) Optical spectrum and (b) AC trace under $I_{SOA}$ = 140 mA and $V_{SA}$ = 0 V. (c) Optical spectrum and (d) AC trace under $I_{SOA}$ = 140 mA and $V_{SA}$ = -1.0 V. (e) Optical spectrum and (f) AC trace under $I_{SOA}$ = 140 mA and $V_{SA}$ = -3.0 V.}\label{fig:5}
\end{figure*}

\begin{figure*}[!htbp]
\centering
  \begin{subfigure}[t]{.16\textwidth}
    \centering
        \captionsetup{justification=centering}
        \caption[c]{}\label{fig:6a}
            \centering
    \includegraphics[width=.99\textwidth]{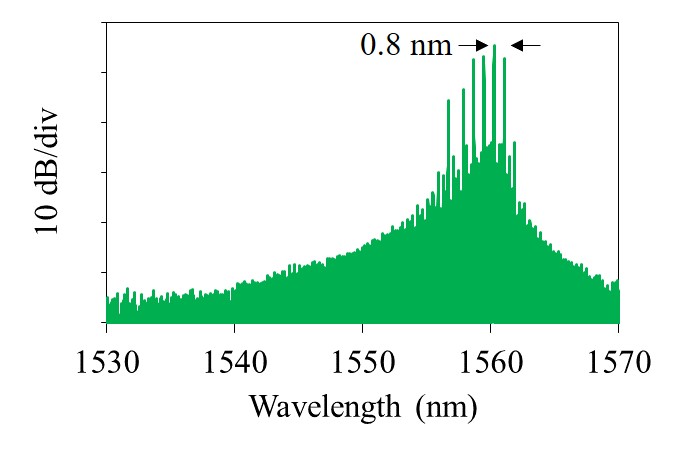}
  \end{subfigure}%
  \begin{subfigure}[t]{.16\textwidth}
    \centering
        \captionsetup{justification=centering}
        \caption[c]{}\label{fig:6b}
            \centering
    \includegraphics[width=.99\textwidth]{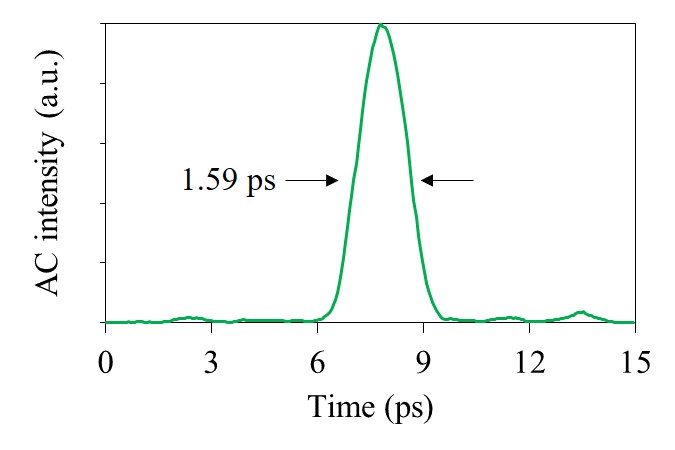}
  \end{subfigure}%
  \begin{subfigure}[t]{.16\textwidth}
    \centering
        \captionsetup{justification=centering}
        \caption[c]{}\label{fig:6d}
            \centering
    \includegraphics[width=.99\textwidth]{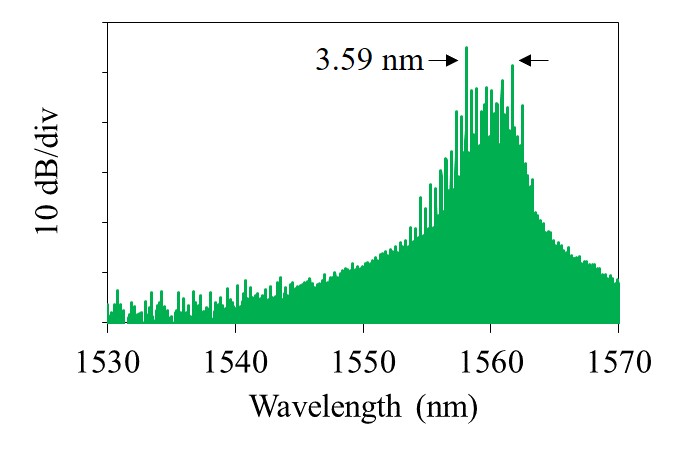}
  \end{subfigure}%
    \begin{subfigure}[t]{.16\textwidth}
    \centering
        \captionsetup{justification=centering}
        \caption[c]{}\label{fig:6e}
            \centering
    \includegraphics[width=.99\textwidth]{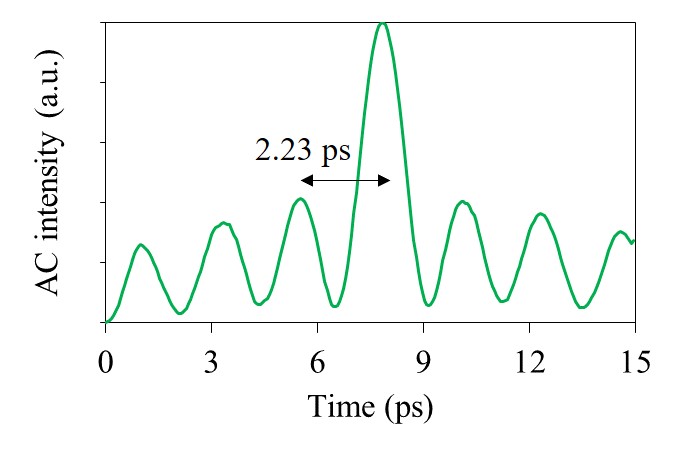}
  \end{subfigure}%
  \begin{subfigure}[t]{.16\textwidth}
    \centering
        \captionsetup{justification=centering}
        \caption{}\label{fig:6f}
            \centering
    \includegraphics[width=.99\textwidth]{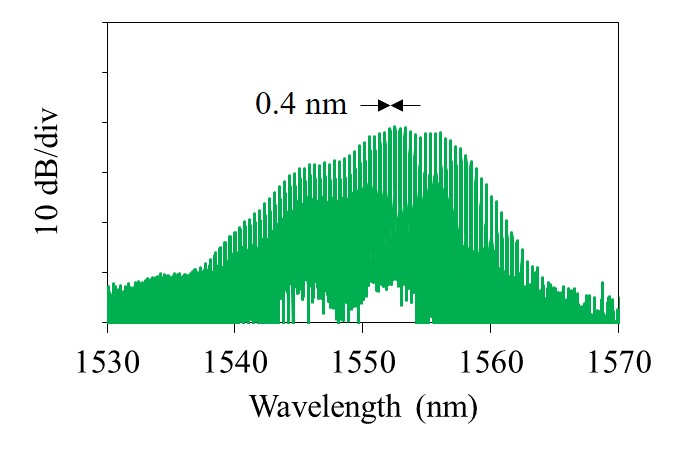}
  \end{subfigure}%
  \begin{subfigure}[t]{.16\textwidth}
    \centering
        \captionsetup{justification=centering}
        \caption{}\label{fig:6h}
            \centering
    \includegraphics[width=.99\textwidth]{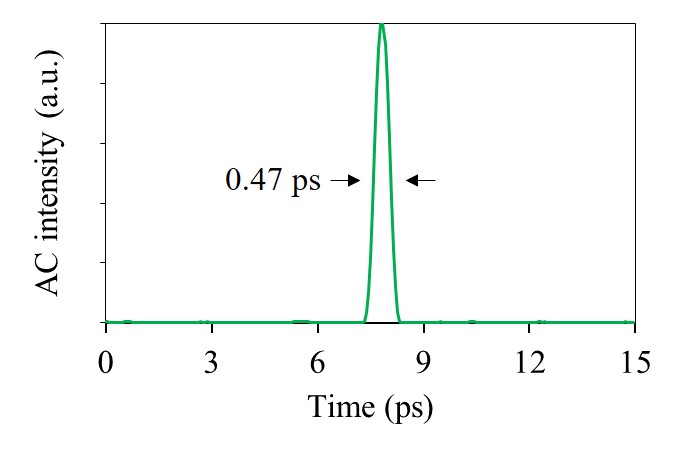}
  \end{subfigure}%
  \caption{a) Optical spectrum and (b) AC trace under $I_{SOA}$ = 150 mA and $V_{SA}$ = 0 V. (c) Optical spectrum and (d) AC trace under $I_{SOA}$ = 150 mA and $V_{SA}$ = -1.0 V. (e) Optical spectrum and (f) AC trace under $I_{SOA}$ = 150 mA and $V_{SA}$ = -3.0 V.}\label{fig:6}
\end{figure*}

As shown in Fig. 3(a), the PIC was mounted on a copper chuck, TEC-controlled at \SI{16}{\celsius} with Thorlabs LD driver, and tested with Yokogawa optical spectrum analyzer (OSA) and APE pulseCheck autocorrelator (AC), following Amonics EDFA with >1-mW output power to meet the sensitivity of AC. In Fig. 3(e), the two L-I curves present the dependency of fiber coupled power on bias current for gain sections, at two different temperatures. 50-mA threshold current and 1.1-mW/A slope efficiency are exhibited at \SI{16}{\celsius}, and 70-mA threshold current and 0.6-mW/A slope efficiency are exhibited at \SI{25}{\celsius}. The I-V curve exhibits a typical semiconductor laser behavior that turning point is below 10 mA, and the voltage constantly rises as bias current increases.

Fig. 3(b) presents the optical spectrum taken under $I_{SOA}$ = 100 mA, $V_{SA}$ = -1.0 V. The optical frequency comb exhibits five comb lines on a left-tilted base, with a 3-dB spectral bandwidth of 1.6 nm and 25.8-dB SMSR (side-mode-suppression-ratio) measured with respect to the fundamental cavity modes.
Fig. 3(c) shows the detailed optical spectrum around 1552 nm, where one can see the modes are suppressed except one out of every four. Thus the mode spacing is 0.8 nm, corresponding to 100 GHz. That is the fourth harmonic of subcavity round-trip frequency 25 GHz. The high-order harmonic spacing (100 GHz) is indicating multiple pulsation in each subcavity round trip (25 GHz), which might be linked to the faster recovery time of saturable absorption usually realized at a higher reverse-bias voltage. 
Fig. 3(d) shows the AC trace under the same bias condition to confirm the mode locking regime. However, the AC trace exhibits a triangular shape rather than the commonly observed Gaussian or squared hyperbolic secant ($Sech^2$). The FWHM (full-width-at-half-maximum) of AC trace is 3.20 ps, and the estimated pulse duration is 2.21 ps using triangular deconvolution factor = 0.692.

Fig. 3(f) presents the optical spectrum taken under same current $I_{SOA}$ = 100 mA but higher reverse-bias voltage $V_{SA}$ = -2.8 V. The wider comb spans over 6.7 nm ($\sim$840 GHz) at 3-dB level, with 22.3-dB SMSR.
Fig. 3(g) shows the detailed optical spectrum around 1552 nm. Every other mode is suppressed, and thus the mode spacing is 0.4 nm corresponding to 50 GHz, the second harmonic of subcavity round-trip frequency (25 GHz). At this bias condition the colliding pulse mode locking regime in the lower subcavity is dominant.
Fig. 3(h) depicts the AC trace. The FWHM of this AC trace is 1.20 ps. Assuming a hyperbolic secant shape the deconvoluted pulse duration is 0.68 ps, in the sub-picosecond region. The time-bandwidth product (TBP) is 0.57, approximately twice as large as the TBP for $Sech^2$-shaped pulses 0.32. The generated pulses might be chirped and could be further compressed. The EDFA might also have contributed dispersion that could be cured by using dispersion compensating fibers (DCF).

Fig. 4 presents a series of optical spectrum and AC trace under $I_{SOA}$ = 140 mA, and $V_{SA}$ = 0, -1.0, and -3.0 V. 
In Fig. 4(a), under $V_{SA}$ = 0 V, the optical mode spacing is 0.8 nm ($\sim$100 GHz), four times the subcavity round-trip frequency (25 GHz). The comb is strongly asymmetric on a left-tilted base.  
Fig. 4(b) presents the triangular AC trace with a FWHM of 3.46 ps. Again, the deconvoluted pulse duration is 2.40 assuming an triangle-shaped pulse. One can hardly see a clear zero-floor, and on both the wings one can even see the falling/rising edge belonging to adjacent pulse peaks.
Fig. 4(c) presents the spectrum under $V_{SA}$ = -1.0 V, two stronger modes separated by 2.79 nm ($\sim$350 GHz) appear that agrees with the temporal displacement of beat-note AC trace in Fig. 4(d), being 2.83 ps ($\sim$350 GHz). The sinusoidal AC trace does not exhibit an equal maximum height at each peak, indicating the retrieved signal may travel in a slowly varying envelope. Frequency purity could be improved by enhancing the suppression ratio of optical frequency comb.
As shown in Fig. 4(e) the mode spacing of the flattened comb under $V_{SA}$ = -3.0 V is 0.4 nm ($\sim$50 GHz). The AC (see Fig. 4(f)) exhibit a narrow coherence spike on clear zero-floor due to a sufficient number of phase-locked modes yielding a stable mode locking. The homogeneity of coherence over such a wide spectral range could be further evaluated by using tunable optical filter. The AC FWHM of 0.67 ps is equivalent to pulse duration of 0.44 ps assuming a $Sech^2$ shape.

Similarly, Fig. 5 presents the optical spectra and AC traces under same voltages $V_{SA}$ = 0, -1.0, and -3.0 V, but little higher gain section bias current $I_{SOA}$ = 150 mA. 
In Fig. 5(a) under $V_{SA}$ = 0 the mode spacing is 0.8 nm ($\sim$100 GHz) and the AC trace has FWHM of 1.59 ps (see Fig. 5(b)). 
In Fig. 5(c) under $V_{SA}$ = -1.0 V, the optical spectrum becomes chaotic but two higher modes with a separation of 3.59 nm ($\sim$450 GHz) give rise to the temporal displacement of 2.23 ps (beat-note frequency of $\sim$450 GHz) of AC trace in Fig. 5(d). The insufficient suppression ratio of optical modes is indicative of spectral impurity in agreement with the slowly varying envelope in AC trace. The nonuniform AC suggests that the deconvolution does not exhibit same intensity, might be linked to the power imbalance between the dual modes. 
In Fig. 5(e) under $V_{SA}$ = -3.0 V, the mode spacing is always 0.4 nm ($\sim$50 GHz). The FWHM of AC is 0.47 ps, corresponding to pulse duration of 0.27 ps, as shown in Fig. 5(f) with a perfectly zero floor. The background-free AC trace resulted from the overall average pulse length proves the stability of pulse train generation. 


The monolithic mode-locked semiconductor laser in a compound ring composed of two coupled linear subcavities has been demonstrated. The compound ring configuration contributes not only optical feedback but also further collisions. The dynamic of collisions and recombination of pulses in each round of splitting into the two linear subcavities might be relevant to the high-order harmonic generation, which is still under theoretic investigation. The experimentally demonstrated optical spectra and AC traces have confirmed the repetition-rate (order of harmonic) switchability from 50 GHz (2nd harmonic) to 450 (18th harmonic) GHz, during the transition between normal passive mode locking and dual-wavelength optical mixing (beat-note). Furthermore, ultrafast pulses down to 0.27 ps have been obtained which might result from improved stability and dispersion engineering that the feedback mechanism provides.

\section{Funding}

European Union’s Horizon 2020 research and innovation programme under the Marie Sklodowska-Curie grant agreement No. 642355 FiWiN5G, Spanish Ministerio de Economia y Competitividad DiDACTIC project (TEC2013-47753-C3-3-R), and Consejería de Educación, Juventud y Deporte of Comunidad de Madrid DIFRAGEOS project (P2013/ICE-3004).

\section{Acknowledgment}

We thank Dr. Julien Javaloyes for fruitful discussions. Also, we acknowledge the support from SMART Photonics B.V (www.smartphotonics.nl), the foundry where the chips were fabricated.

\section{References}



\bigskip

\bibliography{bibi}

\bibliographyfullrefs{bibi}

\end{document}